\documentclass[aps,floats,twocolumn,showpacs]{revtex4-1}
\usepackage{amssymb}
\usepackage{amsmath}
\usepackage{epstopdf}
\usepackage{graphicx}

 \def\ds{\displaystyle}
 \def\bc{\begin{center}}          \def\ec{\end{center}}
 \newcommand{\I}[1]{}
 \newcommand{\angstrom}{\text{\normalfont\AA}}
 \allowdisplaybreaks
\begin{document}
 \title{Long-term evolution of broken wakefields in finite radius plasmas}
 \author{K.V.Lotov}
 \affiliation{Budker Institute of Nuclear Physics SB RAS, 630090, Novosibirsk, Russia}
 \affiliation{Novosibirsk State University, 630090, Novosibirsk, Russia}
 \author{A.V.Petrenko}
 \affiliation{Budker Institute of Nuclear Physics SB RAS, 630090, Novosibirsk, Russia}
 \affiliation{CERN, Geneva, Switzerland}
 \author{A.P.Sosedkin}
 \affiliation{Budker Institute of Nuclear Physics SB RAS, 630090, Novosibirsk, Russia}
 \affiliation{Novosibirsk State University, 630090, Novosibirsk, Russia}
 \date{\today}
 \begin{abstract}
A novel effect of fast heating and charging a finite-radius plasma is discovered in the context of plasma wakefield acceleration. As the plasma wave breaks, the most of its energy is transferred to plasma electrons which create strong charge-separation electric field and azimuthal magnetic field around the plasma. The slowly varying field structure is preserved for hundreds of wakefield periods and contains (together with hot electrons) up to 80\% of the initial wakefield energy.
 \end{abstract}
 \pacs{41.75.Lx, 52.40.Mj, 52.50.Gj}
 \maketitle

Acceleration of charged particles in plasmas, or plasma wakefield acceleration, is currently a hot field of research \cite{RMP85-1,PoP19-055501,RMP81-1229,PoP14-055501}. A plasma can stand electric fields orders of magnitude stronger than those available in metallic or dielectric structures, and this ability is already well proven experimentally \cite{Nat.445-741,NatPhys2-696}. The strength of the electric field in plasma waves is typically limited by the so-called wavebreaking field $E_0 = mc\omega_p/e$, where $m$ is the electron mass, $c$ is the light velocity, $e$ is the elementary charge, and $\omega_p = \sqrt{4 \pi n_0 e^2/m}$ is the plasma frequency determined by the plasma density $n_0$. In other words, the energy density of the plasma wave ($\sim E_0^2/8\pi$) can be as high as the rest energy of plasma electrons.

The great majority of investigations is concentrated at either creation or immediate use of this huge energy density. Long-term evolution of plasma waves at this parameter range has not received much attention. Available studies are focused at dynamics of plasma ions \cite{PoP7-375,PRL86-3332,PRE65-036401,PoP10-1124,PRL109-145005}, turbulization of the wave \cite{PRL86-3332,PoP10-1124,LPB25-313}, creation of soliton-like structures \cite{PRL76-3562,PRL82-3440,PRL83-3434} or generation of the magnetic field \cite{PRL76-2495}. Here we report on the novel aspect of plasma wake behavior which is related to fast charging of the plasma column and  heating of plasma electrons.

The effect was discovered in the context of the AWAKE project \cite{AWAKE,IPAC13-1179,TDR} aimed at the first experimental demonstration of the proton-driven plasma wakefield acceleration \cite{NatPhys9-363,PRST-AB13-041301}. To excite the wakefield efficiently, an initially long proton bunch is transformed into a train of short equally spaced micro-bunches by the self-modulation instability \cite{EPAC98-806,PPCF53-014003,PRL104-255003}. A small precursor is needed to seed the proper instability mode and to speed up growth of the self-modulation \cite{PRST-AB16-041301,PRE86-026402,PoP20-056704}. In the AWAKE the instability is seeded and controlled by the co-propagating laser pulse which instantly creates the plasma by quick ionization of highly uniform \cite{PoP20-013102} neutral gas. If properly injected \cite{PRL107-145003,JPP78-455} into the grown wakefield, witness electrons are accelerated with a sub-GeV/m rate to the energy of about 2 GeV. With the use of higher energy drivers and fine control of the plasma density profile \cite{PoP18-024501}, this method is capable of producing multi-TeV electrons in a single accelerating stage \cite{PoP18-103101}.

The essence of the heating and charging effect is the following. The plasma wave driven by the train of micro-bunches breaks soon after reaching the maximum field amplitude of about $0.4\,E_0$ \cite{PoP20-083119}. The released wave energy is quickly converted into kinetic energy of plasma electrons. The electrons gain high transverse momenta and escape from the wakefield region radially. Since the plasma is created by the short laser pulse, it has a finite radius of several $c/\omega_p$ and a sharp boundary \cite{Oz}. Thus there are not enough cold electrons to replace hot electrons escaping the plasma, at least for the time needed for ionizing the surrounding gas. The uncompensated charge of plasma ions creates the strong radial electric field which in turn keeps most of hot electrons near the plasma.

The escaping electrons predominantly move forward in the laboratory frame and take away some negative current from the plasma. The compensating positive current appears in the plasma and creates a strong azimuthal magnetic field around the plasma.

The effect has an intimate connection with target normal sheath acceleration (TNSA) mechanism of ion acceleration \cite{RMP85-751,RPP75-056401}. The difference is in the geometric configuration of fields and in presence of plasma waves as an intermediate energy carrier.

\begin{figure}[thb]
\bc\includegraphics[width=185bp]{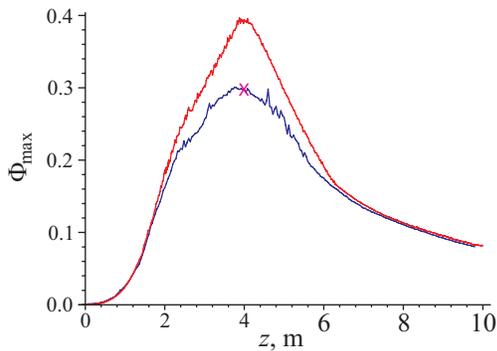} \ec
\caption{Simulated maximum wakefield amplitude versus the propagation distance for infinite (red) and finite-radius (blue) plasmas for parameters of the AWAKE experiments.}\label{f1-growth}
\end{figure}
To illustrate the effect, we use the quasi-static axisymmetric 2d3v code LCODE with the kinetic solver for both plasma electrons and ions \cite{PoP5-785,PRST-AB6-061301,LCODE}. We use cylindrical coordinates $(r, \varphi, z)$ with the $z$-axis as the direction of beam propagation and the time $\tau = t - z/c$ measured from the onset of wave excitation.
The beam and plasma parameters correspond to the baseline parameter set of the AWAKE experiment \cite{TDR,IPAC13-1179}. Here the 400\,GeV proton beam initially focused to 0.2\,mm must propagate through the highly uniform rubidium plasma of length $L_\text{max}=10$\,m and density $n_0 = 7 \times 10^{14}\text{cm}^{-3}$. Along the first 4~meters the beam fully self-modulates; along the following 6~meters the beam creates strong wakefields and accelerates test electrons. The expected plasma radius \cite{Oz} can be approximated as
\begin{equation}\label{e1a}
    r_p = 5 c/\omega_p (1.5 - 0.5\,z/L_\text{max}).
\end{equation}
As a measure of the wakefield amplitude, we take local $\Phi_\text{m} (z,t)$ and absolute $\Phi_\text{max} (z)$ extrema of the dimensionless wakefield potential $\Phi (z,t)$ on the axis:
\begin{equation}\label{e1}
    \Phi (z,t) = \frac{\omega_p}{E_0} \int_{-\infty}^t E_z(z, t') \, d t',
\end{equation}
where $E_z$ is the on-axis electric field. The wakefield potential is more noise-resistant than $E_z$ and contains information on focusing properties of plasma waves in an easy-to-view form. Though the studied examples are based on realistic experimental setups, we will formulate the results in the dimensionless form wherever reasonable.

\begin{figure}[thb]
\bc\includegraphics[width=236bp]{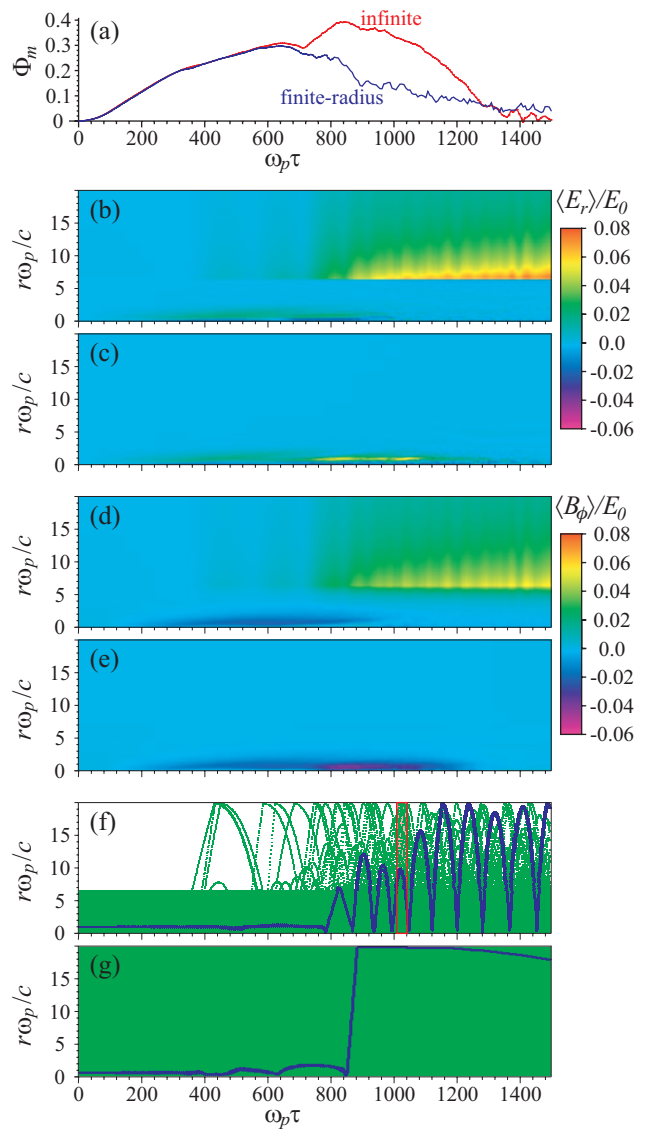} \ec
 \vspace*{-5mm}
\caption{ Time dependence of the wakefield amplitude for finite-radius and infinite plasmas (a), spatial distribution of period-averaged radial electric field $\langle E_r \rangle$ (b,c) and azimuthal magnetic field $\langle B_\phi \rangle$ (d,e), trajectories of selected plasma electrons in the co-moving window (f,g) at the stage of fully modulated beam (at $z=4$\,m) for finite-radius (b,d,f) and infinite (c,e,g) plasmas. In pictures (f,g) one trajectory is shown in blue, while the others are shown in green. The red rectangle in (f) shows the area characterized in Fig.\,\protect\ref{f3-energies}}\label{f2-main}
\end{figure}

First we note that at these parameters neither the finite plasma radius $r_p$ itself, nor its linear variation with length have any significant effect on the amplitude of the excited wave (Fig.{\,}\ref{f1-growth}). The visible difference of maximum amplitudes is due to specific characters of wave breaking near the axis and does not result from different conditions of wave excitation [Fig.{\,}\ref{f2-main}(a)]. Plasma charging is best observed at the time of strongest driver modulation. This place is marked by the cross in Fig.{\,}\ref{f1-growth}.

From the time dependence of the wakefield potential amplitude at $z=4$\,m [Fig.{\,}\ref{f2-main}(a)] we see that the wave breaks at $\tau \approx 700\,\omega_p^{-1}$ in both finite-radius and infinite plasmas. Soon after that, slowly varying radial electric field $E_r$ and azimuthal magnetic field $B_\phi$ appear in the finite-radius case [Fig.{\,}\ref{f2-main}(b,d)]. These fields reach maximum values near the plasma boundary. To separate them from the oscillating wakefields, we average the actual fields over one plasma period ($2 \pi \omega_p^{-1}$). In the infinite plasma, no strong average fields appear [Fig.{\,}\ref{f2-main}(c,e)].

\begin{figure}[tb]
\bc\includegraphics[width=201bp]{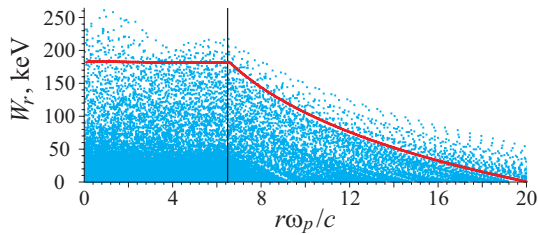} \ec
 \vspace*{-5mm}
\caption{Energy of plasma electrons versus their radial position (dots) and the radial dependence of the electrostatic potential $W_r (r)$ (red line) at the stage of established equilibrium ($\tau=1120\,\omega_p^{-1}$, $z=4$\,m). The black vertical line shows the plasma boundary.}\label{f3-energies}
\end{figure}
Trajectories of plasma electrons responsible for charging the plasma are shown in Fig.{\,}\ref{f2-main}(f). The high-energy electrons appear at the time of wave breaking. First of them dart directly to the outer wall located at $r_\text{max}=20\,c/\omega_p$ and carry out the negative charge. Later produced fast electrons are confined by the generated electric field, but still make long radial excursions. In the code, as soon as an electron hits the wall it is reflected back inelastically with some low energy. This models emission of secondary electrons. The secondary electrons are pulled back to the plasma and also participate in formation of the electron halo between the plasma and the wall. In the infinite plasma, high energy electrons are also generated as the wave breaks [Fig.{\,}\ref{f2-main}(g)], but the bulk of cold electrons fully compensate the escaping charge. Correspondingly, no secondary electrons are pulled back toward the plasma axis.

Figure~\ref{f3-energies} shows the energy distribution of the halo electrons at the stage of established equilibrium and the shape of the electrostatic potential
\begin{equation}\label{e3}
    W_r (r) = e \int_r^{r_\text{max}} \langle E_r \rangle dr,
\end{equation}
where the angle brackets denote time averaging over the interval $2 \pi \omega_p^{-1}$. We see that both the typical electron energy and the depth of the potential well for them are about 200\,keV. This corresponds to the electric field as strong as 200 MV/m at the plasma boundary.

\begin{figure}[b]
\bc\includegraphics[width=235bp]{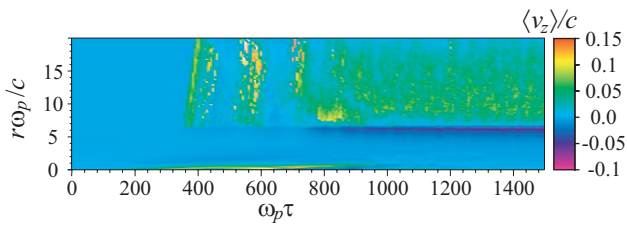} \ec
 \vspace*{-5mm}
\caption{Period-averaged velocity of plasma electrons at $z=4$\,m.}\label{f4-vz}
\end{figure}
The average longitudinal motion of plasma electrons is best viewed on the map of electron velocities (Fig.\,\ref{f4-vz}). The electron halo predominantly moves forward, while the compensating current is concentrated in the area of width $\sim c/\omega_p$ near the plasma boundary. Correspondingly, the magnetic field penetrates the outer plasma region [Fig.\,\ref{f2-main}(d)] unlike the electric field which is sharply screened by the surface charge [Fig.\,\ref{f2-main}(b)]. This field behavior resembles penetration of slowly-varying electromagnetic wave into the plasma in classical electrodynamics.

\begin{figure}[htb]
\bc\includegraphics[width=206bp]{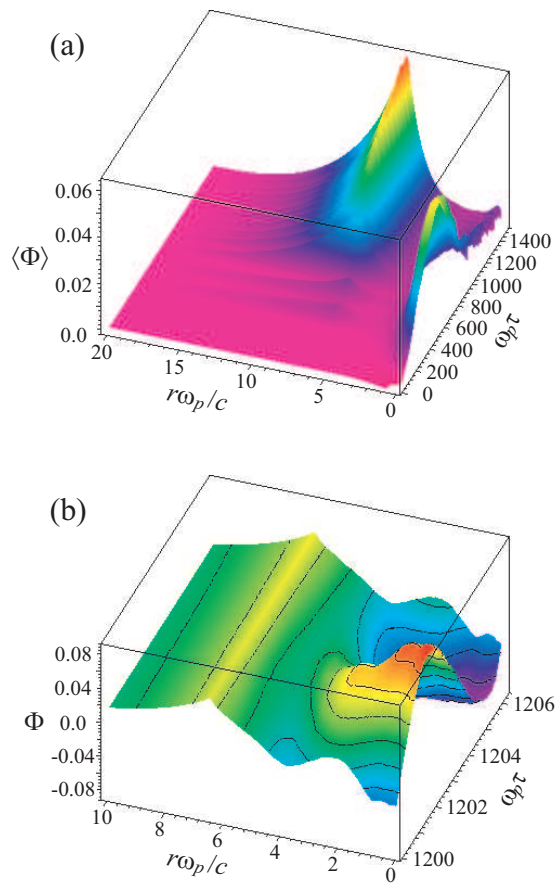} \ec
 \vspace*{-5mm}
\caption{The wakefield potential at $z=4$\,m: period-averaged slowly evolving part (a) and one-period fragment showing the oscillating part (b).}\label{f5-phi}
\end{figure}
The overall action of electric and magnetic fields on axially moving relativistic particles is characterized by the wakefield potential $\Phi$ (Fig.\,\ref{f5-phi}), which has an off-axis slowly-varying spike, i.e., the
potential well for electrons. The spike is located exactly at the plasma boundary, so its location can be controlled by changing the plasma radius. The height of the off-axis spike is comparable with on-axis potential oscillations caused by residual plasma wakefields [Fig.\,\ref{f5-phi}(b)]. However, the energy stored in slowly varying fields is much greater than that in plasma oscillations due to the geometrical factor (since the off-axis area is larger).

The energy balance in the system can be quantitatively described by the energy flux $\Psi$ in the light-velocity frame \cite{PRE69-046405}:
\begin{multline}\label{e4}
    \Psi = \int_0^\infty \biggl[ \sum_j m_j c^2 (\gamma_j-1) (c-v_{j,z}) \\
    + \frac{c}{8\pi}(E^2+B^2) - \frac{c}{4 \pi} (E_r B_\varphi - E_\varphi B_r) \biggr] 2 \pi r \, dr,
\end{multline}
where $m_j$, $\vec v_j$, and $\gamma_j$ are mass, velocity, and relativistic factor of individual plasma particles; the sum is over plasma particles in a unit volume. The energy flux $\Psi$ changes as the current $j_{bz}$ of the relativistic beam works against the wakefield:
\begin{equation}\label{e5}
    \frac{\partial \Psi}{\partial \tau} = - c \int_0^\infty j_{bz} E_z 2 \pi r \, dr - c Q,
\end{equation}
where $Q$ is the power absorbed by a unit length of the wall. Analyzing different terms in (\ref{e4}) allows us to separate the energy accumulated in particles and fields. We can find that in the finite-radius plasma 82\% of the energy left by the driver remains in the plasma: 17\% in the fields and 65\% in the motion of plasma electrons. In contrast, in the infinite plasma about 15\% of the energy remains in the plasma after wave breaking and only 5\% is the field energy. The latter number gives us the estimate for the energy remained in residual wakefields ($\sim 10$\%), as the field energy is approximately the half of the plasma wave energy.

The lifetime of the formed field structure can be limited either by ion motion or by ionization of the surrounding gas. The time it takes for the electric field to displace a plasma ion of the mass $M_i$ is of the order of
\begin{equation}\label{e6a}
\tau_\text{ion} = \omega_p^{-1} \sqrt{M_i E_0 /(m \langle E_r \rangle)}
\end{equation}
or 1.2\,nsec (300 wave periods) for the considered plasma and $\langle E_r \rangle = 0.05\,E_0$. After that time, ion motion comes into play and takes the energy from fields and hot electrons in a way similar to TNSA mechanism.

\begin{figure}[t]
\bc\includegraphics[width=200bp]{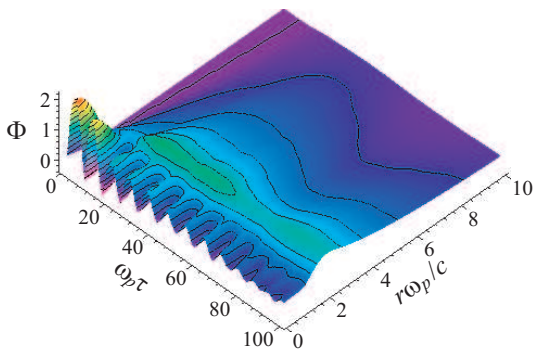} \ec
 \vspace*{-5mm}
\caption{The wakefield potential produced by the short laser pulse.}\label{f6-laser}
\end{figure}
The ionization time is determined by the cross-section of electron impact ionization. For the lower estimate of the electron energy (1\,keV), typical values of ionization cross-section $\sigma_i$ range from $0.15\,\angstrom^2$ for hydrogen and helium to several $\angstrom^2$ for heavier elements \cite{r2}. For the rubidium plasma ($\sigma_i \approx 1.5~\angstrom^2$, \cite{r3}), this corresponds to the mean free path $\lambda_i = 1/(n\sigma_i) \sim 10$\,cm and ionization time $\sim 5$\,nsec. Since for non-relativistic electrons with energies $W_E\gg 100$\,eV the ionization cross-section scales as $W_E^{-1}$ \cite{r1}, and most outer electrons have energies well above 1\,keV (Fig.\,\ref{f3-energies}), the average ionization time is even longer and much longer than the ion response time. Thus the lifetime of the slowly varying fields is determined by ion motion and is as long as hundreds of plasma periods.

The physical effects responsible for formation of slowly varying electromagnetic fields and the cloud of hot electrons are not specific to proton drivers and can be also reproduced with a simpler setup. To illustrate this, we simulate long-term evolution of the wakefields created by a short laser pulse in a hydrogen plasma of the radius $3.5\,c/\omega_p$. The pulse shape is defined by the normalized vector potential squared as
\begin{gather}\label{e6}
    |a|^2 = \frac{\ds a_0^2 \, e^{-r^2/\sigma_r^2}}{2} \left[
  1 + \cos \left(\frac{\sqrt{\pi} c (\tau -\tau_c)}{\sigma_z \sqrt{2}} \right) \right],
\\ \nonumber |\tau -\tau_c| < \sigma_z \sqrt{2 \pi} / c
\end{gather}
with $a_0 = 5$, $\sigma_r = c/\omega_p$, $\sigma_z = \sqrt{2} c/\omega_p$, and $\tau_c = 3.55\,\omega_p^{-1}$. The wall is located at $r_\text{max} = 40\,c/\omega_p$.

All the discussed above features are reproduced with this driver, namely: a slowly varying radial electric field around the plasma ($\sim 0.5\,E_0$), a slowly varying azimuthal magnetic field penetrating outer plasma layers (also $\sim 0.5\,E_0$), a smooth hill of the wakefield potential (Fig.\,\ref{f6-laser}), and a cloud of hot electrons around the plasma. The energy balance here is the following: 80\% of the energy delivered by the driver remains with the plasma, 13\% remains in the electromagnetic fields and 67\% is the kinetic energy of plasma particles.

To conclude, we discovered that the energy stored in the plasma in the form of a high-amplitude plasma wave is eventually converted to the thermal energy of a small fraction of plasma electrons. If the plasma has a finite radius, then the most of this energy remains near the plasma, since fast electrons create strong charge-separation electric field and are kept by this field around plasma ions. Due to specific nature of wave breaking, there also appears a strong azimuthal magnetic field around the plasma. The tubular potential well (or hill) associated with these fields may find use for focusing, concentrating, or scattering relativistic particles. The location of the potential spike can be controlled by changing the plasma radius, as the two coincide. Perhaps, further optimization for a particular need may result in even stronger or smoother fields or higher efficiency of energy transfer from the drive beam to the slowly varying field structure.

This work was supported by The Ministry of education and science of Russia, Siberian Supercomputer Center SB RAS, and RFBR grant 14-02-00294.


\begin{thebibliography}{18}
 \bibitem{RMP85-1}
    S. Corde, K. Ta Phuoc, G. Lambert, R. Fitour, V. Malka, A. Rousse, A. Beck, and E. Lefebvre,
    Rev. Mod. Phys. {\bf 85}, 1 (2013).
 \bibitem{PoP19-055501}
    V.Malka,
    Phys. Plasmas {\bf 19}, 055501 (2012).
  \bibitem{RMP81-1229}
   E. Esarey, C. B. Schroeder, and W. P. Leemans,
   Rev. Mod. Phys. \textbf{81}, 1229 (2009).
  \bibitem{PoP14-055501}
	C. Joshi,
   	Phys. Plasmas, v.14 (2007), p.055501.
  \bibitem{Nat.445-741}
	I. Blumenfeld, C.E. Clayton, F.-J. Decker, M.J. Hogan, C. Huang, R. Ischebeck, R. Iverson, C. Joshi, T. Katsouleas, N. Kirby, W. Lu, K.A. Marsh, W.B. Mori, P. Muggli, E. Oz, R.H. Siemann, D. Walz, and M. Zhou,
	Nature {\bf 445}, 741 (2007).
  \bibitem{NatPhys2-696}
W.P.Leemans, B.Nagler, A.J.Gonsalves, Cs.Toth, K.Nakamura, C.G.R.Geddes,
E.Esarey, C.B.Schroeder, and S.M.Hooker, GeV electron beams from a
centimetre-scale accelerator.
    Nature Physics, v.2 (2006), p.696--699.
\bibitem{PoP7-375}
    L.M.Gorbunov, P.Mora, R.R.Ramazashvili, A.A.Solodov,
    Phys. Plasmas \textbf{7}(1), 375 (2000).
\bibitem{PRL86-3332}
    L.M.Gorbunov, P.Mora, and A.A.Solodov,
    Phys. Rev. Lett. \textbf{86}, 3332-3335 (2001).
\bibitem{PRE65-036401}
    L.M.Gorbunov, P.Mora, and R.R.Ramazashvili,
    Phys. Rev. E \textbf{65}, 036401 (2002).
  \bibitem{PoP10-1124}
    L.M.Gorbunov, P.Mora, and A.A.Solodov.
    Phys. Plasmas, v.10 (2003), ¹4, p.1124--1134.
\bibitem{PRL109-145005}
    J.Vieira, R.A.Fonseca, W.B.Mori, and L.O.Silva,
    Phys. Rev. Lett. \textbf{109}, 145005 (2012).
\bibitem{LPB25-313}
    C.T.Zhou, M.Y.Yu, and X.T.He,
    Laser and Particle Beams \textbf{25}, 313 (2007).
\bibitem{PRL76-3562}
    S.V.Bulanov, M.Lontano, T.Zh.Esirkepov, F.Pegoraro, A.M.Pukhov.
    Phys. Rev. Lett. \textbf{76}, 3562 (1996).
\bibitem{PRL82-3440}
    S.V.Bulanov, T.Zh.Esirkepov, N.M.Naumova, F.Pegoraro, and V.A.Vshivkov,
    Phys. Rev. Lett. \textbf{82}, 3440-3443 (1999).
\bibitem{PRL83-3434}
    Y.Sentoku, T.Zh.Esirkepov, K.Mima, K.Nishihara, F.Califano, F.Pegoraro, H.Sakagami, Y.Kitagawa,
N.M.Naumova, and S.V.Bulanov,
    Phys. Rev. Lett. \textbf{83}, 3434-3437 (1999).
\bibitem{PRL76-2495}
    L.Gorbunov, P.Mora, and T.M.Antonsen, Jr.,
    Phys. Rev. Lett. \textbf{76}, 2495 (1996).
\bibitem{AWAKE}
    AWAKE Collaboration,
    Proton-driven plasma wakefield acceleration: a path to the future of high-energy particle physics
    (submitted to Plasma Physics and Controlled Fusion, 2014).
 \bibitem{IPAC13-1179}
	P. Muggli, A. Caldwell, O. Reimann, E. Oz, R. Tarkeshian, C. Bracco, E. Gschwendtner, A. Pardons, K. Lotov, A. Pukhov, M. Wing, S. Mandry, J. Vieira,
	Physics of the AWAKE Project.
	Proceedings of IPAC2013 (Shanghai, China), p.1179-1181.
 \bibitem{TDR}
	AWAKE Collaboration,
	AWAKE Design Report: A Proton-Driven Plasma Wakefield Acceleration Experiment at CERN.
	CERN-SPSC-2013-013; SPSC-TDR-003.
 \bibitem{NatPhys9-363}
    A.Caldwell, K.Lotov, A.Pukhov, and F.Simon,
    Nature Phys. {\bf 5}, 363 (2009).
 \bibitem{PRST-AB13-041301}
	K.V.Lotov,
	Phys. Rev. ST Accel. Beams {\bf 13}, 041301 (2010).
 \bibitem{EPAC98-806}
    K.V.Lotov,
    Instability of long driving beams in plasma wakefield accelerators.
    Proc. 6th European Particle Accelerator Conference (Stockholm, 1998), p.806-808.
 \bibitem{PPCF53-014003}
	A. Caldwell, K. Lotov, A. Pukhov and G. Xia,
	Plasma Phys. Controlled Fusion {\bf 53}, 014003 (2011).
 \bibitem{PRL104-255003}
	N.Kumar, A.Pukhov, and K.Lotov,
	Phys. Rev. Lett. {\bf 104}, 255003 (2010).
 \bibitem{PRST-AB16-041301}
	K.V.Lotov, G.Z.Lotova, V.I.Lotov, A.Upadhyay, T.Tuckmantel, A.Pukhov, A.Caldwell,
	Phys. Rev. ST Accel. Beams 16, 041301 (2013).
\bibitem{PRE86-026402}
    C.B.Schroeder, C.Benedetti, E.Esarey, F.J.Gruner, and W.P.Leemans,
    Phys. Rev. E \textbf{86}, 026402 (2012).
\bibitem{PoP20-056704}
    C.B.Schroeder, C.Benedetti, E.Esarey, F.J.Gruner, and W.P.Leemans,
    Phys. Plasmas \textbf{20}, 056704 (2013).
 \bibitem{PoP20-013102}
	K.V.Lotov, A.Pukhov, and A.Caldwell,
	Phys. Plasmas 20(1), 013102 (2013).
 \bibitem{PRL107-145003}
	A. Pukhov, N. Kumar, T. Tuckmantel, A. Upadhyay, K. Lotov, P. Muggli, V. Khudik, C. Siemon, and G. Shvets,
	Phys. Rev. Lett. 107(14), 145003 (2011).
 \bibitem{JPP78-455}
	K.V.Lotov,
	J. Plasma Phys. 78(4), 455-459 (2012).
 \bibitem{PoP18-024501}
	K.V.Lotov,
	Phys. Plasmas 18(2) (2011) 024501.
 \bibitem{PoP18-103101}
	A. Caldwell  and K. V. Lotov,
	Phys. Plasmas {\bf 18}, 103101 (2011).
 \bibitem{PoP20-083119}
    K.V.Lotov,
    Phys. Plasmas 20, 083119 (2013).
\bibitem{Oz}
    E.Oz, P.Muggli,
    A Novel Rb Vapor Plasma Source for Plasma Wakefield Accelerators.
    (to appear in Nucl. Instr. Meth. A, 2014).
\bibitem{RMP85-751}
    A.Macchi, M.Borghesi, M.Passoni,
    Rev. Mod. Phys. \textbf{85}, 751 (2013).
\bibitem{RPP75-056401}
    H.Daido, M.Nishiuchi, A.S.Pirozhkov,
    Rep. Prog. Phys. \textbf{75} (2012) 056401.
\bibitem{PoP5-785}
    K.V.Lotov,
    Phys. Plasmas, 1998, v.5, N 3, p.785-791.
\bibitem{PRST-AB6-061301}
    K.V.Lotov,
    Phys. Rev. ST Accel. Beams {\bf 6}, 061301 (2003).
\bibitem{LCODE}
 \verb"www.inp.nsk.su/~lotov/lcode".
 \bibitem{PRE69-046405}
    K.V.Lotov,
    Phys. Rev. E, v.69 (2004), N 4, p.046405(1-13).
\bibitem{r2}
    P. L. Bartlett and A. T. Stelbovics. Phys. Rev. A \textbf{66}, 012707 (2002).
\bibitem{r3}
    Yong-Ki Kim. J. Migdalek, W. Siegel, J. Bieron. Phys. Rev. A \textbf{57}, 246 (1998).
\bibitem{r1}
    L. D. Landau, L. M. Lifshitz. Quantum Mechanics: Non-Relativistic Theory, 3rd Edition, Elsevier, 2003.

\end{thebibliography}
\end{document}